 \pdfoutput=1
\documentclass{ifacconf}

\usepackage{amssymb,mathtools,mathrsfs,bm}
\usepackage{array, multirow}
\usepackage{tabularx, graphics, graphicx,subfigure}
\usepackage{natbib}
\usepackage{booktabs} % special rule in table
\begin{document}
\begin{frontmatter}

\title{Improved Pump Setpoint Selection Using a Calibrated Hydraulic Model of a High-Pressure Irrigation System} 
% Title, preferably not more than 10 words.

% \thanks[footnoteinfo]{Sponsor and financial support acknowledgment
% goes here. Paper titles should be written in uppercase and lowercase
% letters, not all uppercase.}

\author[unimelb-EEE]{Ye Wang}
\author[unimelb-IE]{Qi Zhao}
\author[unimelb-IE,uniad]{Wenyan Wu}
\author[lmw]{Ailsa Willis}
\author[uniad,unimelb-IE]{Angus R. Simpson} 
\author[unimelb-EEE]{Erik Weyer}

\address[unimelb-EEE]{Department of Electrical and Electronic Engineering, The University of Melbourne, Victoria 3010, Australia\\ (E-mail: \{ye.wang1, ewey\}@ unimelb.edu.au)}
\address[unimelb-IE]{Department of Infrastructure Engineering, The University of Melbourne, Victoria 3010, Australia\\ (E-mail: \{jason.zhao1, wenyan.wu\}@ unimelb.edu.au)}
\address[uniad]{School of Civil, Environmental \& Mining Engineering, The University of Adelaide, South Australia 5005, Australia\\ (E-mail: angus.simpson@adelaide.edu.au)}
\address[lmw]{Lower Murray Water, Mildura, Victoria 3500, Australia \\(E-mail: ailsa.willis@lmw.vic.gov.au)}

\begin{abstract}               
    % Abstract of not more than 250 words.
    This paper presents a case study of the operational management of the Robinvale high-pressure piped irrigation water delivery system (RVHPS) in Australia. Based on datasets available, improved pump setpoint selection using a calibrated hydraulic model is investigated. The first step was to implement pre-processing of measured flow and pressure data to identify errors in the data and possible faulty sensors. An EPANET hydraulic simulation model was updated with calibrated pipe roughness height values by using the processed pressure and flow data. Then, new pump setpoints were selected using the calibrated model given the actual measured demands such that the pressures in the network were minimized subject to required customer service standards. Based on a two-day simulation, it was estimated that 4.7\% savings in pumping energy cost as well as 4.7\% reduction in greenhouse gas emissions can be achieved by applying the new pump setpoints.
\end{abstract}


\end{frontmatter}
%===============================================================================

\section{Introduction}

Water distribution systems are critical infrastructure that supply water to municipal (industrial, commercial and residential) and rural (irrigation and residential) users \citep{cantoni2007control,gu2020irrigation,wang2017non}. The growth in water demand has led to larger energy consumption and costs, as well as larger greenhouse gas (GHG) emissions (when fossil fuel sources are used), typically due to pumping \citep{wu2020changing}. As a leader in climate change mitigation in the state of Victoria in Australia, the water sector has committed to reducing its emissions by 42\% by 2025 and to net-zero emissions by 2050, under Victoria’s water plan, Water for Victoria \citep{envreport}. A reduction of energy consumption is therefore of critical importance and it also brings economic and environmental benefits.

Robinvale High Pressure System (RVHPS) is one of four irrigation systems managed by Lower Murray Urban and Rural Water Corporation (LMW) \citep{lmwreport1,lmwreport2}. The network has been pipelined over time, initially with gravity spurs supplied from a main channel. The operation of RVHPS is energy intensive. In the last four full financial years, RVHPS incurred approximately 35\% of electricity cost for irrigation assets. Most of consumed energy is taken from the grid while a on-site solar energy production contributes to approximately 3\% total energy consumption of RVHPS. During peak demand seasons, the RVHPS pump station will often run at close to full capacity for 24 hours a day to meet both irrigation demands and the pressure head requirement. 

This paper gives a description of RVHPS and available data, which enables an understanding of the system characteristics and allows an evaluation of potential cost savings and GHG emission reductions. Using real data from RVHPS, data pre-processing of flow and pressure data has been implemented to detect and identify possible errors in data as well as faulty sensors. Then, using processed flow and pressure data, a hydraulic model of RVHPS was calibrated (using an optimization technique) by estimating the best pipe roughness height values. The calibrated hydraulic model was then used to analyze the system hydraulics as well as to develop an improved control strategy. In addition, the service requirement in the form of minimum pressure head at all the irrigation outlets was considered in the development of an improved pump setpoint selection algorithm.

%%%%%%%%%%%%%%%%%%%%%%%%%%%%%%%%%%%%%%%%%%%%%%%%%%%%
%%%%%%%%%%%%%%%%%%%%%%%%%%%%%%%%%%%%%%%%%%%%%%%%%%%%
\section{Description of the Robinvale High-Pressure Irrigation System}\label{section:description}

\begin{figure}[thbp]
    \begin{center}
        \includegraphics[width=0.8\hsize]{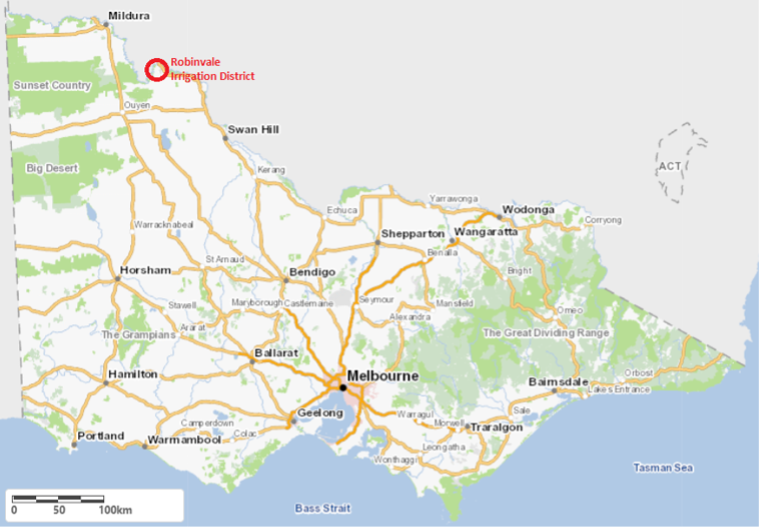}    % The printed column width is 8.4 cm.
        \caption{Location of the Robinvale Irrigation District.} 
        \label{fig:location}
    \end{center}
\end{figure}

\begin{figure}[thbp]
    \begin{center}
        \includegraphics[width=0.8\hsize]{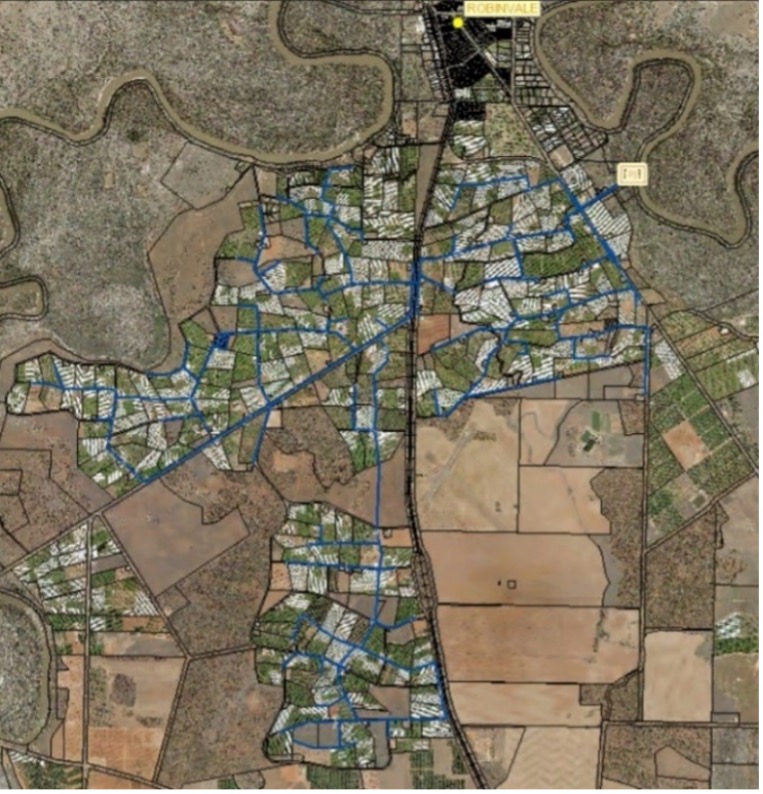}    % The printed column width is 8.4 cm.
        \caption{Aerial View of the Robinvale Irrigation District.} 
        \label{fig:aerial view}
    \end{center}
\end{figure}

The RVHPS is located in the Robinvale Irrigation District. The District is situated on the south bank of the Murray River in north-western Victoria \citep{lmwreport1}, as shown in Fig.~\ref{fig:location}. The irrigation district covers an area of approximately 2,700 hectares. Water is pumped from the Murray River and delivered to customers for both irrigation and domestic and stock (D\&S) use. Table grapes are the major crop planted in this area \citep{lmwreport2}, which requires large volumes of water for irrigation. Water use for crop cooling creates additional demand, particularly during the early afternoon (12 pm to 6 pm) on high demand days. An aerial view of the irrigation district is shown in Fig. \ref{fig:aerial view}.

The Robinvale irrigation network was fully pipelined with the commissioning of the current RVHPS in October 2010. There are 433 pipes, 244 irrigation outlets and 210 small diameter D\&S outlets in the system. Irrigation water needs to be ordered in advance, while D\&S water can be used from any of the outlets in the network without requiring an order. The maximum flow rate for a D\&S outlet is 0.75 L/s. LMW is currently running a water ordering management tool which allows farmers to order water (for irrigation use) in multiples of an hour in advance. In the peak demand season (summer), some farmers may not be able to order water at their preferred time since the total capacity of the network is limited. Water ordering is used to avoid demand exceeding pumping and network capacity. In general, compliance with orders is important to support sharing of network capacity.

The Robinvale high-pressure pump station has a capacity of about 3700 L/s. A total of 3000 L/s is available for irrigation water orders, 300 L/s is reserved for D\&S use and the remaining 400 L/s accommodates farmers using flow rate in excess of orders, turning an outlet off late or starting an order early. During a hot summer, the pump station will often run at almost full capacity 24 hours a day to meet both the peak demand as well as the minimum service pressure head requirement downstream of irrigation outlets. As a result, the energy cost and the associated GHG emissions can be high during the peak season.

%%%%%%%%%%%%%%%%%%%%%%%%%%%%%%%%%%%%%%%%%%%%%%%%%%%%
%%%%%%%%%%%%%%%%%%%%%%%%%%%%%%%%%%%%%%%%%%%%%%%%%%%%
\section{Hydraulic Model Calibration}\label{section:model calibration}

\subsection{EPANET Hydraulic Model}

\begin{figure}[thbp]
    \begin{center}
        \includegraphics[width=\hsize]{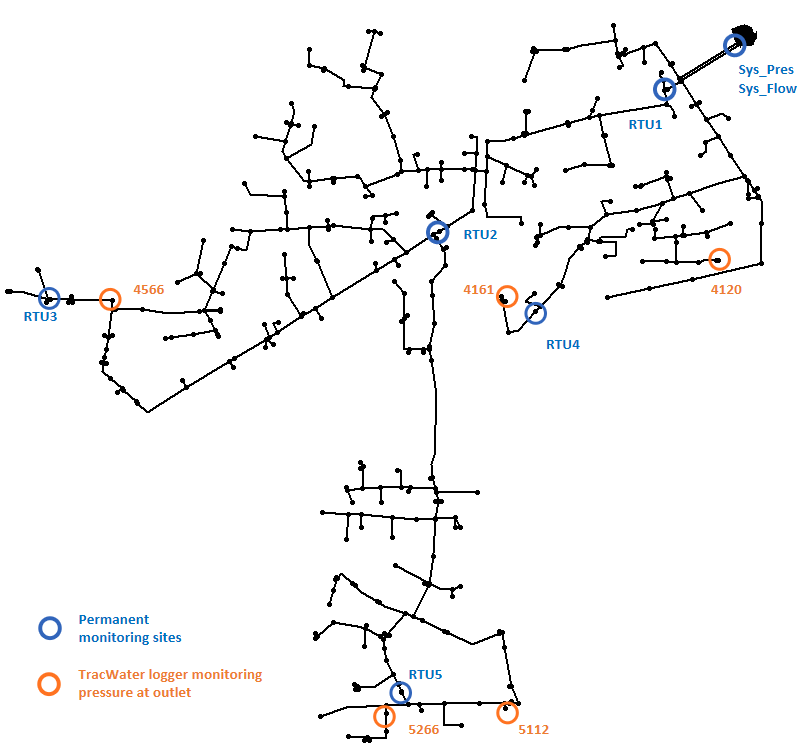}    % The printed column width is 8.4 cm.
        \caption{The EPANET model and Sensor Locations of the Robinvale Irrigation District.} 
        \label{fig:sensors}
    \end{center}
\end{figure}

EPANET is an open-source software package developed to perform extended period simulation of system hydraulics within a pressurised network \citep{epanet}. An EPANET hydraulic model of the RVHPS piped network had already been built by LMW in 2010. As shown in Fig. \ref{fig:sensors}, the EPANET model consists of 435 nodes with the ground elevations at nodes in the network ranging from 47 m to 75 m, and 433 pipes with pipe lengths ranging from 0.7 m to 1359 m and pipe diameters ranging from 225 mm to 1400 mm.

\subsection{Available Data and Data Pre-processing}

In the Robinvale irrigation system, each irrigation outlet has a remotely monitored electromagnetic flow meter so that near real-time demand data is available. As shown in Fig. \ref{fig:sensors}, there are six permanent pressure sensors installed in the system. One pressure sensor (Sys$_{-}$Pres) is located at the pump station and the other five (RTU1 to RTU5) across the network. In addition, permanent flow meters (Sys$_{-}$Flow) are installed at the pump station, measuring the total flow into the piped network. All pressure and flow data at permanent monitoring sites are regularly collected by a SCADA system.

Pump station pressure and flow data is monitored via an Ethernet cable connection and recorded at 2-second intervals. Outlet real-time demands are checked at five-minute intervals, and flow rate is recorded if a change exceeding a deadband value has occurred. Pressure at five irrigation outlets is recorded at 15-minute intervals using temporary local data loggers and pressure sensors. Pressure at the five network sensors is polled at regular intervals. All data has been interpolated or down sampled to a 1-min sampling time interval.

Through a few initial runs of the EPANET model simulation, large differences between the field-observed and model-simulated pressures were identified at pressure sensors: RTU2, RTU3 and RTU5. At RTU2, the observed hydraulic grade line (HGL) was constantly approximately 10 m higher than the system HGL, which indicated the need for error investigation. Theoretically, under a zero-flow condition, the difference between pressures at two nodes is expected to be equal to the difference between the elevations of the two nodes. After making the comparison under several historical static-pressure (near-zero flow) conditions, constant monitoring errors were identified to be 0.2 m, -12.6 m, 3.0 m, 0.9 m and 3.5 m for RTU1 to RTU5, respectively. In addition, as the sensor elevations were checked and confirmed to be accurate, the corresponding errors were corrected at the five RTU locations.

\subsection{Hydraulic Model Calibration}

In this study, pipe roughness height values were calibrated for the RVHPS hydraulic model. In order to eliminate the impact of pump operations on the calibration results and mainly focus on the targeted pipe roughness, we consider the observed system pressures downstream of the pump station.

For the RVHPS, the flows in each pipe and the heads at each node are estimated in the EPANET model using the continuity equations and the pipe head loss equations, which can be formulated as follows \citep{todini1988gradient}:
\begin{subequations}\label{eq:continuity and headloss}
    \begin{align}
        & \mathbf{A}_1^{\top} \mathbf{q}(t) + \mathbf{d}(t) = 0,\\
        & - \mathbf{G}(t) \mathbf{q}(t) + \mathbf{A}_1 \mathbf{h}(t) + \mathbf{A}_2 \mathbf{e} = 0,
    \end{align}
\end{subequations}
where $\mathbf{q}(t) = [q_1(t),\ldots,q_P(t)]^\top $ denotes the vector of pipe flows in $P$ pipes, $\mathbf{h}(t) = [h_1(t),\ldots,h_N(t)]^\top $ denotes the vector of heads at $N$ nodes at time $t$, $\mathbf{d}(t) = [d_1(t),\ldots,d_N(t)]^\top$ denotes the vector of demands for $N$ demand nodes at time $t$, $\mathbf{e} = [e_{1},\ldots,e_{F}]^\top $ denotes the vector of elevations for $F$ fixed-head nodes, $\mathbf{A}_1$ is the link node incidence matrix of $P$ pipes by $N$ nodes and $\mathbf{A}_2$ is the reservoir node incidence matrix of $P$ pipes by $F$ fixed-head nodes. Furthermore, the matrix $\mathbf{G}(t)$ can be defined based on the Darcy-Weisbach friction loss formula as
\begin{align}\label{eq:resistance}
    \mathbf{G}(t) = \begin{bmatrix}
        r_1(\varepsilon_1) |q_1(t)| & \cdots & 0\\
        \vdots & \ddots & \vdots\\
        0 & \cdots & r_P(\varepsilon_P) |q_P(t)|
    \end{bmatrix},
\end{align}
where $\varepsilon_i$, $i=1,\ldots,P$ are the pipe roughness height values for each pipe. The resistance term $r_i(\varepsilon_i)$ can be computed by
\begin{align}
    r_i(\varepsilon_i) = \frac{f_i(\varepsilon_i) L_i}{ 2 g D_i A_i^2}, \; i = 1,\ldots,P,
\end{align}
where $L_i$, $D_i$ and $A_i$ represent the length, diameter, area of the $i$-th pipe, respectively, and $g$ is the gravitational acceleration. The friction factor $f_i(\varepsilon_i)$ is approximated as
\begin{align}\label{eq:friction}
    f(\varepsilon_i) = \frac{0.25}{\Big ( \log_{10} \Big( \frac{\varepsilon_i}{3.7D_i} + \frac{5.74}{\mathbf{Re}^{0.9}}\Big) \Big )^2},
\end{align}
where $\mathbf{Re}$ is the Reynolds number.

The boundary conditions for the pipe head loss equations are given by elevation of fixed nodes (reservoirs and water storage tanks) and also the head range considered from the pump curve.

The calibration of pipe roughness height values can be achieved by solving an optimization problem. In this study, we investigated several different choices of decision variables. One way of grouping pipes is demonstrated in this study, where roughness heights of the four different pipe materials in the system are considered as four decision variables. As pipe material is a key feature in determining pipe roughness, the simplifying assumption is that all pipes of the same material have the same roughness heights. The four types of pipes are MSCL (mild steel cement mortar lined), GRP (glass reinforced plastic), DICL (ductile iron cement mortar lined) and mPVC (modified polyvinyl chloride). MSCL pipes are the two DN1200\footnote{DN1200: nominal diameter 1200 mm} rising mains out of the pump station. GRP pipes are mainly large trunk mains with diameters ranging from DN1000 to DN1400. DICL pipes are mainly smaller distribution mains with diameters ranging from DN375 to DN750. mPVC pipes have the smallest diameters (DN225 to DN375) connected to irrigation outlets.

The overall calibration objective is to minimize the mismatch between the observed and model simulated pressure heads at the five pressure monitoring sites $j=1,\ldots,5 $ (RTU1 to RTU5). Considering a total number of simulation time steps $T>0$, the calibration optimization problem can be formulated as follows:
\begin{equation}\label{problem:calibration}
    \underset{\varepsilon_1,\ldots, \varepsilon_P}{\mathrm{minimize}}\;\; \frac{1}{T} \sum_{t=1}^{T} \sum_{j=1}^{5} \left( p_{o_j}(t) - p_{m_j}(t) \right)^2,
\end{equation}
where $p_{o_j}(t)$ and $p_{m_j}(t)$ are field-observed and model simulated pressure heads at time $t$ at sites RTU1 to RTU5, respectively. The simulated pressure head $p_{m_j}(t)$ can be obtained from solving \eqref{eq:continuity and headloss} with $h_j(t) = p_{m_j}(t) + z_{j}$, $j=1,\ldots,5$, where $z_{j}$ is the ground surface elevation. In this study, we use the EPANET to solve \eqref{eq:continuity and headloss} and obtain the simulated heads $ h_j(t) $ at each time $t$ at the five pressure monitoring sites.

\subsection{Hydraulic Model Calibration Results}

Sequential least squares programming has been used to solve the calibration optimization problem in \eqref{problem:calibration}. The 28th Dec 2019 was selected as the calibration period. The optimal solutions for the calibrated pipe roughness height values for each type of pipe are reported in Table \ref{table:pipe roughness options}. As shown in the results, large roughness values, particularly for MSCL and GRP pipes, have been obtained. This is very likely caused by a significant growth of biofilms in pipes, as the network delivers raw river water that contains a large amount of nutrients. The minimum sum of the RMSEs from the five pressure sensor locations was calculated to be 8.543 m. The breakdown of the average observed and modelled values, percentages, as well as the individual Root Mean Square Error (RMSE) and Mean Absolute Error (MAE) at each monitoring site, is summarized in Tables~\ref{table:calibration results-pressure} and \ref{table:calibration results-flow}. From these two tables, it can be seen that the calibrated EPANET model can simulate the RVHPS and obtain similar pressures and flows as observed in the field.

\begin{table}[thbp]
    \begin{center}
        \caption{Calibrated Pipe Roughness Values}\label{table:pipe roughness options}
        \begin{tabular}{ccccc}
        \toprule
        Pipe Material & MSCL & DICL & GRP & mPVC \\\midrule
        Roughness Height [mm] & 10.6 & 0.44 & 2.91 & 0.01\\
        \bottomrule
        \end{tabular}
    \end{center}
\end{table}

\begin{table*}[thbp]
    \begin{center}
        \caption{Calibration Results at Pressure Monitoring Sites}\label{table:calibration results-pressure}
        \begin{tabular}{cccccc}
        \toprule
        Site & Average Observed Head [m] & Average Simulated Head [m] & Perentage Diff. & RMSE [m] & MAE [m] \\\midrule
        RTU1 & 141.4 & 141.5 & -0.09 \% & 0.340 & 0.234 \\
        RTU2 & 138.6 & 137.1 & 1.03 \% & 1.739 & 1.440 \\
        RTU3 & 132.0 & 133.3 & -1.02 \% & 1.977 & 1.595 \\
        RTU4 & 135.7 & 135.1 & 0.38 \% & 2.515 & 1.876 \\
        RTU5 & 131.7 & 131.8 & -0.05 \% & 1.972 & 1.123 \\
        \bottomrule
        \end{tabular}
    \end{center}
\end{table*}

\begin{table*}[thbp]
    \begin{center}
        \caption{Calibration Results at Flow Monitoring Sites}\label{table:calibration results-flow}
        \begin{tabular}{cccccc}
        \toprule
        Site & Average Observed Flow [L/s] & Average Simulated Flow [L/s] & Perentage Diff. & RMSE [L/s] & MAE [L/s] \\\midrule
        System Flow & 2528 & 2501 & 1.06\% & 52.90 & 26.93\\
        \bottomrule
        \end{tabular}
    \end{center}
\end{table*}

\subsection{Hydraulic Model Validation}

\begin{figure}[thbp]
    \begin{center}
        \subfigure[System Flows]{\includegraphics[width=0.8\hsize]{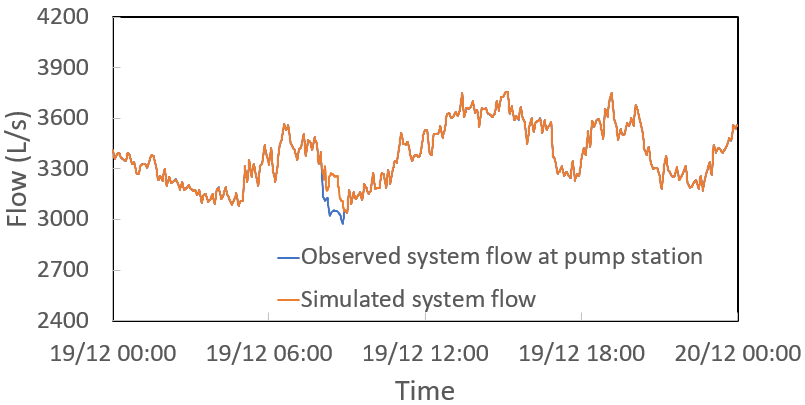}}   % The printed column width is 8.4 cm.
        \subfigure[Pressure Heads at RTU4]{\includegraphics[width=0.8\hsize]{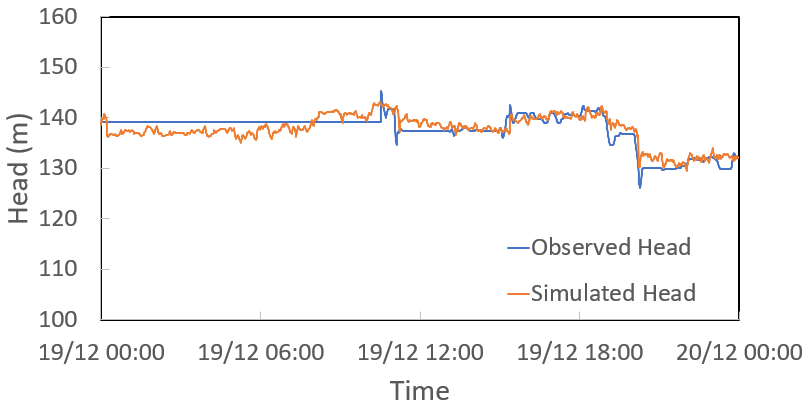}}    % The printed column width is 8.4 cm.
        \caption{Model Validation Results.}
        \label{fig:validation}
    \end{center}
\end{figure}

Model validation is an important step to verify the performance of the calibrated hydraulic model. The validation period was selected to be 19th December 2019, which is one of the days with the highest demands. Selected results are shown in Fig. \ref{fig:validation}. The observed heads are smooth especially before 10 am due to that small changes in pressure were not recorded. The observed and simulated system flows are very similar as well as the heads at RTU4. Thus, it can be concluded that a good match was achieved between the observed flow and pressure data and the simulated data from the calibrated EPANET model.

%%%%%%%%%%%%%%%%%%%%%%%%%%%%%%%%%%%%%%%%%%%%%%%%%%%%
%%%%%%%%%%%%%%%%%%%%%%%%%%%%%%%%%%%%%%%%%%%%%%%%%%%%
\section{Improved Pump Setpoint Selection Algorithm}\label{section:control}

The control objective for the RVHPS is to operate the pump station to provide sufficient water flow with at least a minimum allowable pressure head at the irrigation outlets. In this section, we propose an algorithm to select pressure head setpoint at the pump station, which aims at saving energy as well as guaranteeing the minimum required downstream pressure at all the irrigation outlets.

\subsection{Proportional-Integral Controller}

The current operations of RVHPS pumps are governed by a proportional-integral (PI) controller, which calculates pump speed and the number of pumps operating based on the difference between the desired pressure setpoint $r(t)$ and the pressure head $h_p(t)$ at the exit of the pump station. The coefficients of the PI controller have been set up properly so that any desired pressure setpoint $r(t)$ can be reached in a short transition period.

The desired pressure setpoint $r(t)$ is currently selected by using a pressure setpoint curve, which relates the total system flow to the pressure setpoint. The range of the pressure setpoints is from 81.9 m to 102.3 m and the setpoint selection is based on the system total flow in a range between 0 L/s to 3500 L/s. The system flow is measured in real-time and is fed back to automatically choose a corresponding pressure setpoint from the pressure-setpoint curve.

\subsection{Head Loss Estimation across Irrigation Outlets}

In the current pumping operation, the head loss across each irrigation outlet is not considered. The minimum pressure head delivered downstream of the irrigation outlets should equal or exceed 35 m. The minimum pressure head targeted in the new control strategy therefore equals 35 m plus the total head loss across the irrigation outlet, which includes the head loss through the control valve and other outlet components (pipes, bends, one flow meter and one butterfly valve). Every irrigation outlet in the network incorporates a three-way pilot control valve with pressure reducing, pressure sustaining and flow control pilots. Head loss across the irrigation outlet can be significant. Therefore, it is important to estimate the head loss, which focuses on the minimum pressure at upstream of the exit from each outlet.

\begin{figure}[t]
    \begin{center}
        \includegraphics[width=\hsize]{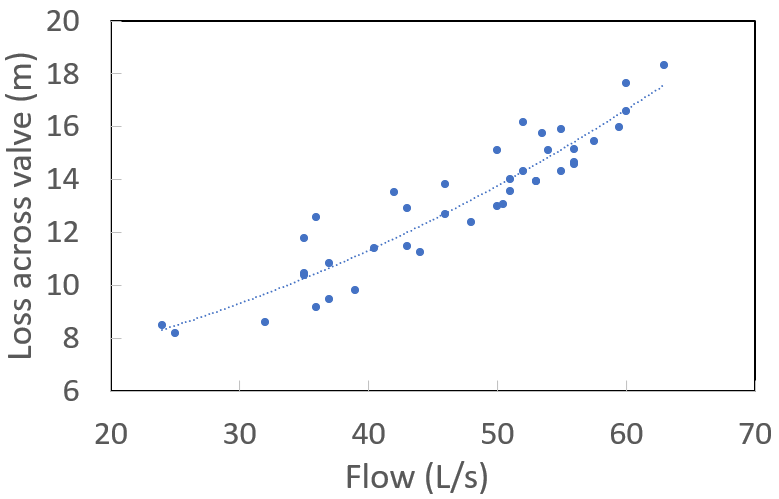}    % The printed column width is 8.4 cm.
        \caption{Head Loss across three DN150 Outlets.} 
        \label{fig:pressure loss}
    \end{center}
\end{figure}

Total head loss across each irrigation outlet is estimated by summing the losses through a valve with control pilots and other outlet components. Head loss across the valve assembly and 0.3 m of pipe was measured in the field for three outlets as shown in Fig. \ref{fig:pressure loss}. Therefore, the head loss across each outlet is estimated by developing a curve relating the pressure drop and flow based on historical pressure observations. We use a second-order polynomial equation to model this head loss as follows:
\begin{align}\label{eq:headloss}
    p_l(q(t)) = a_2 q^2(t) + a_1 q(t) + a_0,
\end{align}
where $a_0 = 5.95 $, $a_1 = 0.0456 $ and $a_2 = 0.00221$. This curve in \eqref{eq:headloss} was used to estimate head losses across each DN150 outlet in the RVHPS.

\subsection{Improved Pump Setpoint Selection}

Based on the available data, any significant improvement in the performance of the PI controller does not seem possible. However, energy savings can be achieved by where possible, by lowering the pump setpoint. 

Considering a 15-minute sampling time interval, improved new pump setpoints may be obtained based on the following steps:
\begin{enumerate}
    \item[(i)] Start with an initial pump pressure setpoint $r(t) = r_0$. 
    \item[(ii)] Run the calibrated EPANET hydraulic simulation model with measured actual irrigation delivery flows $d_i(t)$, $i=1,\ldots, M$ from the SCADA system for one sampling time interval.
    \item[(iii)] Read the EPANET model upstream pressure heads $p_{m_i}(t)$, $i=1,\ldots, M$ for each active irrigation outlet $i$.
    \item[(iv)] Obtain outlet downstream pressure heads $p_{d_i}(t)$, $i=1,\ldots, M$ for each outlet $i$ by
    \begin{align}
        p_{d_i}(t) = p_{m_i}(t) - p_l(d_i(t)),
    \end{align}
    where the head loss across the irrigation outlet $p_l(d_i(t))$ is approximated by \eqref{eq:headloss}.
    \item[(v)] Find the critical pressure head $\underline{p_{d_i}}(t) $ as the minimum pressure head at any active irrigation outlet by
    \begin{align}
        \underline{p_{d_i}}(t) = \min \{p_{d_i}(t), i =1,\ldots, M \},
    \end{align}
    subject to $d_i(t)>0$. The constraint indicates active irrigation outlets currently taking water from the network. 
    \item[(vi)] Adjust the new pump setpoint by
    \begin{align}
        r(t+1) = r(t) + \Delta r(t),
    \end{align}
    with $\Delta r(t) = 35 - \underline{p_{d_i}}(t) $. This correction term $\Delta r(t)$ is chosen based on the HGL of the whole network satisfying \eqref{eq:continuity and headloss}-\eqref{eq:friction} from the EPANET model.
    \item[(vii)] Repeat step (ii) at the next sampling time instant.
\end{enumerate}

The water ordering system ensures that the setpoint is in a feasible range, as it will reject orders that cannot be satisfied. However, in practice, non-compliant water usage behavior can happen at some irrigation outlets. This non-compliant behavior will potentially cause an unsatisfactory level of service at some irrigation outlets.

%%%%%%%%%%%%%%%%%%%%%%%%%%%%%%%%%%%%%%%%%%%%%%%%%%%%
%%%%%%%%%%%%%%%%%%%%%%%%%%%%%%%%%%%%%%%%%%%%%%%%%%%%
\subsection{Control Operation Results}

\subsubsection{Comparison of Pump Pressure Setpoints}

\begin{figure}[b]
    \begin{center}
        \includegraphics[width=\hsize]{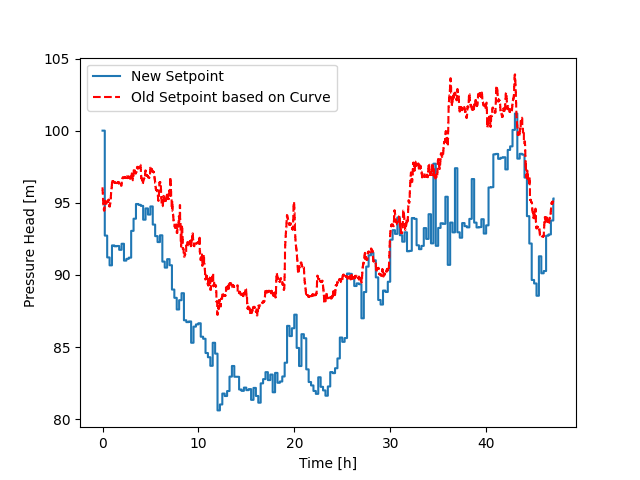}    % The printed column width is 8.4 cm.
        \caption{Comparison of New and Old Pump Setpoints.} 
        \label{fig:setpoint comparison}
    \end{center}
\end{figure}

The economic and environmental benefits that can be achieved using the improved new setpoints are demonstrated using historical data from 28th and 29th Dec 2019. A comparison of new pump setpoints and those obtained using the original pressure setpoint curve is shown in Fig. \ref{fig:setpoint comparison}. The old setpoints are mostly higher than the new setpoints. The average reduction in setpoints values in Fig. \ref{fig:setpoint comparison} is 4.50 m. The operations with the new pump setpoints over the two days investigated results in a 7.08 MWh savings in pumping energy, a \$800 savings in pumping energy cost, and a reduction of 7.72 tonnes in GHG emissions. The operation with the new pump setpoints shows 4.7\% savings in energy consumption and GHG emissions compared to the old setpoints from the curve during peak demand periods. This shows that the new setpoints can lead to lower energy consumption and lower associated GHG emissions related to the system pumping operation while delivering close or equal to the minimum required 35 m pressure head downstream of all active irrigation outlets that actually take water from the network.

\begin{table}[t]
    \begin{center}
        \caption{Comparison Results with Two-day Simulation}\label{table:comparison setpoints}
        \begin{tabular}{ccc}
        \toprule
         & Old & New \\\midrule
        Average Pump Setpoint & 94.0 m & 89.5 m \\
        Total Energy & 149.3 MWh & 142.2 MWh \\
        Total Energy Cost & \$16,300 & \$15,500 \\
        Volume Pumped & 464 ML & 464 ML \\
        Unit Energy Usage & 322 kWh/ML & 306 kWh/ML \\
        GHG Emissions & 163 tonnes & 155 tonnes \\
        \bottomrule
        \end{tabular}
    \end{center}
\end{table}

\subsubsection{Comparison of Level of Service}

\begin{figure}[thbp]
    \begin{center}
        \subfigure[Pump Setpoints]{\includegraphics[width=0.8\hsize]{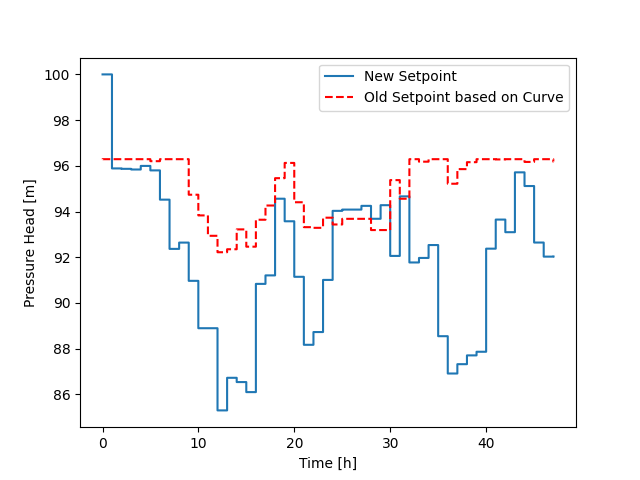}} 
        \subfigure[Measure of Level of Service]{\includegraphics[width=0.8\hsize]{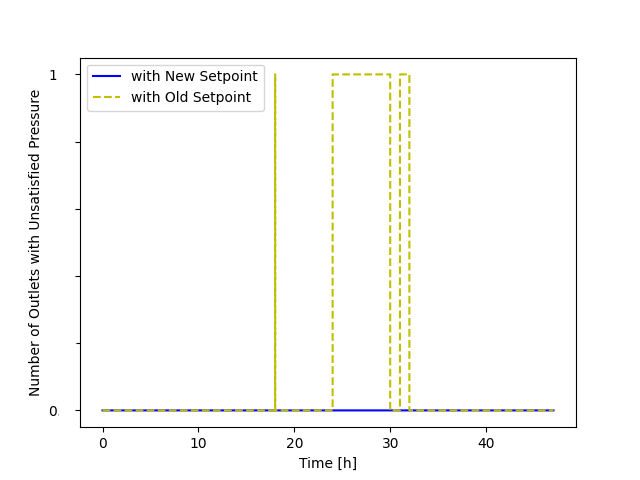}}    % The printed column width is 8.4 cm.
        \caption{Comparison of Level of Service (Minimum pressure head of 35 m delivered at each irrigation outlet).} 
        \label{fig:LoS}
    \end{center}
\end{figure}

A comparison of the level of service has been considered, as shown in Fig.~\ref{fig:LoS}. The case was simulated taking into account that all the farmers comply with their water orders. Higher new setpoints were chosen in order to provide required pressure with compliance to water orders. In this case, operation with the new pump setpoints ensures that the required minimum pressure head (35 m) is delivered at all the irrigation outlets every 15 minutes when the pump setpoint is recalculated. The operations with the old setpoints did not always 
meet pressure requirements at all the irrigation outlets. The largest unsatisfied pressure head magnitude was 3.51 m for a duration of approximately 3 hours.

%%%%%%%%%%%%%%%%%%%%%%%%%%%%%%%%%%%%%%%%%%%%%%%%%%%%
%%%%%%%%%%%%%%%%%%%%%%%%%%%%%%%%%%%%%%%%%%%%%%%%%%%%
\section{Conclusions}\label{section:conclusion}

We have presented a case study of the RVHPS in this paper. Using available measured real-time operational data, an EPANET hydraulic model of the RVHPS was calibrated to update pipe roughness height values. The calibrated EPANET hydraulic model was subsequently used in the development of an improved pump pressure setpoint selection algorithm. The proposed algorithm makes sure the minimum required downstream pressure at all the irrigation outlets is met when customers comply with water orders. Through two-day simulation results, it was found that the improved pump pressure setpoints leads to a 4.7\% reduced energy costs and GHG emissions reduction and an increased level of service to farmers.

\begin{ack}
    The authors would like to thank the LMW for providing operational and design data of the RVHPS, and also thank the Faculty of Engineering and Information Technology (FEIT), the University of Melbourne for funding through the 2020 interdisciplinary grant program. Ye Wang and Wenyan Wu acknowledge support from the Australian Research Council via the Discovery Early Career Researcher Awards (DE220100609 and DE210100117), respectively.
\end{ack}

\bibliography{ifacconf}             % bib file to produce the bibliography
                                                     % with bibtex (preferred)
                                                   
%\begin{thebibliography}{xx}  % you can also add the bibliography by hand

%\bibitem[Able(1956)]{Abl:56}
%B.C. Able.
%\newblock Nucleic acid content of microscope.
%\newblock \emph{Nature}, 135:\penalty0 7--9, 1956.

%\bibitem[Able et~al.(1954)Able, Tagg, and Rush]{AbTaRu:54}
%B.C. Able, R.A. Tagg, and M.~Rush.
%\newblock Enzyme-catalyzed cellular transanimations.
%\newblock In A.F. Round, editor, \emph{Advances in Enzymology}, volume~2, pages
%  125--247. Academic Press, New York, 3rd edition, 1954.

%\bibitem[Keohane(1958)]{Keo:58}
%R.~Keohane.
%\newblock \emph{Power and Interdependence: World Politics in Transitions}.
%\newblock Little, Brown \& Co., Boston, 1958.

%\bibitem[Powers(1985)]{Pow:85}
%T.~Powers.
%\newblock Is there a way out?
%\newblock \emph{Harpers}, pages 35--47, June 1985.

%\bibitem[Soukhanov(1992)]{Heritage:92}
%A.~H. Soukhanov, editor.
%\newblock \emph{{The American Heritage. Dictionary of the American Language}}.
%\newblock Houghton Mifflin Company, 1992.

%\end{thebibliography}

% \appendix
% \section{A summary of Latin grammar}    % Each appendix must have a short title.
% \section{Some Latin vocabulary}              % Sections and subsections are supported  
                                                                         % in the appendices.
\end{document}